\documentclass[aps, pre, twocolumn, superscriptaddress, nofootinbib, amsmath, amssymb]{revtex4-1}

\usepackage{graphicx}  
\usepackage{mathrsfs} 
\usepackage[normalem]{ulem} 

\begin{document}

\title{Topological critical slowing down:\\ 
variations on a toy model
}

\author{Claudio Bonati}
\email{claudio.bonati@df.unipi.it}
\affiliation{Dipartimento di Fisica e Astronomia dell'Universit\`a di
  Firenze and \\ INFN - Sezione di Firenze, Via Sansone 1, 50019, Sesto
  Fiorentino (FI), Italy.}
\altaffiliation{Present address: Dipartimento di Fisica dell'Universit\`a di Pisa and INFN
  - Sezione di Pisa, Largo Pontecorvo 3, I-56127 Pisa, Italy.}

\author{Massimo D'Elia}
\email{massimo.delia@unipi.it}
\affiliation{Dipartimento di Fisica dell'Universit\`a di Pisa and INFN
  - Sezione di Pisa,\\ Largo Pontecorvo 3, I-56127 Pisa, Italy.}

\begin{abstract}
Numerical simulations of lattice quantum field theories whose continuum
counterparts possess classical solutions with non-trivial topology face a
severe critical slowing down as the continuum limit is approached. Standard
Monte-Carlo algorithms develop a loss of ergodicity, with the system remaining
frozen in configurations with fixed topology. We analyze the problem in a
simple toy model, consisting of the path integral formulation of a quantum
mechanical particle constrained to move on a circumference. More specifically,
we implement for this toy model various techniques which have been proposed to
solve or alleviate the problem for more complex systems, like non-abelian gauge
theories, and compare them both in the regime of low temperature and in that of
very high temperature. Among the various techniques, we consider also a new
algorithm which completely solves the freezing problem, but unfortunately is
specifically tailored for this particular model and not easily exportable to
more complex systems.
\end{abstract}

\maketitle

\section{Introduction}

Numerical Monte-Carlo simulations of lattice gauge theories are currently one
of the most important tools for the non-perturbative study of quantum field
theories for fundamental interactions and condensed matter systems.  The
general idea of the method is to reduce the computation of the path integral to
the sampling of a finite dimensional probability distribution, a task that can
be performed through Markov chain Monte-Carlo (MCMC).

In some cases, some hard problems are found which prevent a full exploitation
of numerical simulations. A particularly severe problem that is sometimes
encountered is the loss of the positivity of the probability distribution, a
fact generically indicated as ``sign problem''. This happens for the cases of
QCD at finite baryon density or in the presence of a $\theta$ term and in many
condensed matter models \cite{Aarts:2015tyj, Loh90, Troyer:2004ge}.  In some
other cases the problem is simply the presence of long autocorrelation times,
related to some particular slow modes, leading eventually to a loss of
ergodicity and thus to a breakdown of the algorithm. Such a behaviour is common
to many complex systems and it appears, for instance, close to a phase
transition \cite{Berg_book, NB_book}.

A well known problem of the second type is related to the presence of classical
solutions with non-trivial topology. In the continuum, the space of field
configurations contributing to the path integral divides in homotopy classes,
each class being characterized by the value of a topological invariant taking
only discrete (typically integer) values. On a discrete space-time the concept
of homotopy, which strictly speaking is lost, is recovered as the continuum
limit is approached. Since the typical MCMC algorithms move in the
configuration space in an approximately continuous way, close to the continuum
they become completely unable to change the topology of the field.
 
This problem severely affects the study of $\theta$-dependence in QCD
\cite{Alles:1996vn, Schaefer:2010hu, Bonati:2015vqz} and in many QCD-like
models \cite{Campostrini:1992ar, DelDebbio:2002xa, DelDebbio:2004xh,
Flynn:2015uma}, the most relevant case being the study of the axion potential
at finite temperature \cite{Berkowitz:2015aua, Borsanyi:2015cka,
Kitano:2015fla, Trunin:2015yda, Bonati:2015vqz, Petreczky:2016vrs,
Frison:2016vuc, Borsanyi:2016ksw}.  Several methods and new algorithms have
been proposed to solve or at least alleviate this problem, however no
comparative investigation of the efficiency of these proposals exists in the
literature so far. The purpose of this study is to make a critical comparison
of the various techniques in one of the simplest model where an analogous
problem appears: the numerical simulation of the thermal path integral of a
quantum particle constrained to move on a circumference; we will consider both
the free case and the case in which a potential is present (quantum pendulum).

The use of such a simple model will enable us to extract the autocorrelation
time of the topological susceptibility for several values of the lattice
spacing and to study its critical behaviour as the continuum limit is
approached.  This behaviour (e.g. exponential or power-law in $1/a$) is
expected to be characteristic of the updating scheme, independently of the
specific model adopted as far as topologically stable classical solutions
exist. Let us stress that, instead, the prefactor in front of the leading
$a$-dependence, which eventually fixes what is the best algorithm to be chosen
for a given value of $a$, is likely dependent on the chosen model. However our
main focus here is on the approach to the continuum limit: that is pre-factor
independent and contains relevant information which is expected to be model
independent.

We will show that all methods proposed so far to sample the configuration space
present some residual critical slowing down as the continuum limit is
approached, even if some of them reduce autocorrelation times by several orders
of magnitude with respect to standard local algorithms.  We will also present a
new algorithm, which is based on the idea that the correct way to deal with the
problem is to invent some non-local move in configuration space, capable of
jumping directly from one topological sector into another one. Such a move can
be easily found in the simple model studied in this paper and completely kills
the critical slowing down of the topological modes. The basic idea of
this update scheme seems to be applicable also to more interesting theories,
like 4d non-abelian gauge theories, however its actual implementation requires
further studies to overcome some technical difficulties.

The paper is organized as follows: in Sec.~\ref{sec:model} we introduce the
model and we review some basic facts about the $\theta$ dependence, while in
Sec.~\ref{sec:disc} the discretization to be used in the simulations is
described. With Sec.~\ref{sec:metro} we start the analysis of the different
algorithms that can be used to numerically investigate the topological
properties of the model at $T=0$, providing results for the autocorrelation
times obtained by using the different approaches (standard Metropolis, tailor
method, slab method, open boundary conditions, parallel tempering). In
Sec.~\ref{sec:highT} we study instead the high-temperature limit of the model,
that presents peculiar difficulties. Finally in Sec.~\ref{sec:concl} we present
our conclusions.

\section{The model and some of its properties}\label{sec:model}

The fact that a simple unidimensional quantum mechanical model can have some
common features with four-dimensional non-abelian gauge theories can be at
a first sight quite surprising, however (as far as the $\theta$ dependence is
concerned) what matters is topology: for a quantum particle moving on the
circumference, the topology of the configuration space, $\pi_1(S^1)=\mathbb{Z}$;
for a 4d non-abelian quantum field theory, the topology of the gauge group,
$\pi_3(SU(N))=\mathbb{Z}$.

Let us consider the $\theta = 0$ case first; here and in the following we set
$\hbar=1$.  The thermal partition function $Z$ of a quantum free particle of
mass $m$, moving along a circumference of radius $R$ at temperature
$k_BT=1/\beta$, can be written as a sum over energy (angular momentum)
eigenstates 
\begin{equation}
Z = \sum_{n = -\infty}^{\infty}
\exp\left( -\beta \frac{n^2}{2 m R^2} \right)
\end{equation}
while in the path integral approach
\begin{equation}
\begin{aligned}
&Z = \int\mathscr{D} x(\tau) \exp \left(- S_E[x(\tau)]\right)\ , \\
&S_E = \int_0^{\beta} \mathrm{d} \tau\  \frac{1}{2} m 
\left( \frac{\mathrm{d}x}{\mathrm{d} \tau}\right)^2\ ,
\end{aligned}
\end{equation}
where $S_E$ is the Euclidean action, the path integral extends over periodic
paths, $x(0)=x(\beta)$, and is characterized by the fact that only continuous
paths, even if not differentiable, contribute to it.  For that reason,
analogously to what happens for the configurations contributing to the path
integral of $SU(N)$ non-abelian gauge theories, paths divide in homotopy
classes (topological sectors), classified by a topological number $Q \in \pi_1
(S^1) = \mathbb{Z}$, which corresponds to the number of times the path winds
around the circumference while winding around the Euclidean time circle.

Further insight in the origin of these common features, including
the introduction of the $\theta$ parameter, is obtained by means of
the canonical quantization approach \cite{Jackiw:1979ur, CurrAlg_book}.  The
Hamiltonian $\hat{H}$ of a quantum particle moving on a circumference commutes
with the unitary operator $\hat{U}$ that implements the translation $\phi\to
\phi+2\pi$, where $\phi$ is the angle that parametrizes the position on the
circumference.  As a consequence we can use a base of common eigenvectors of
$\hat{H}$ and $\hat{U}$ and restrict ourselves to the subspace corresponding to
the eigenvalue $e^{i\theta}$ of $\hat{U}$.  The same argument can be applied to
the case of 4d non-abelian gauge theories: local gauge invariance, in the form
of the Gauss law, implies that local gauge transformations act trivially on the
whole Hilbert space, but the theory is invariant also under transformations
having nontrivial behaviour at infinity.  The operator that implements these
``large'' gauge transformations is the analogous of the $\hat{U}$ operator in
the quantum mechanics example (for more details see, e.g., Ref.~\cite{Smilga_book,
StrocchiQFT_book}).

To see that the restriction to the $\theta$-sector of the Hilbert space is
equivalent to the introduction of a $\theta$-term in the Lagrangian it is
convenient to use the path integral formulation.  We can use the identity
$\sum_{Q=-\infty}^{+\infty}e^{-iQ(\theta-\theta')}=2\pi\delta(\theta-\theta')$
to fix the constraint $\hat{U}\psi(\phi, t)=e^{i\theta}\psi(\phi, t)$, thus
obtaining
\begin{widetext}
\begin{equation}
\begin{aligned}
{}_{\theta}\langle \phi_f,t_f| \phi_i, t_i\rangle_{\theta}&=
\frac{1}{2\pi}\sum_{Q\in\mathbb{Z}} e^{-iQ\theta}
\int_{\phi(t_i)=\phi_i}^{\phi(t_f)=\phi_f+2\pi Q} \mathscr{D} \phi(t) 
\exp\left(i\int_{t_i}^{t_f}L(\phi)\mathrm{d} t\right)=\\
&=\frac{1}{2\pi}\int_{\phi(t_i)=\phi_i}^{\phi(t_f)=\phi_f} \mathscr{D} \phi(t)
\exp\left(i\int_{t_i}^{t_f}\left(L(\phi)-\frac{\theta}{2\pi}\dot{\phi}\right)
\mathrm{d} t\right)\ ,
\end{aligned}
\end{equation}
\end{widetext}
where $L(\phi)$ is the original unconstrained Lagrangian.

The restriction can be interpreted also as a modification of the properties of
the physical states for translations $\phi \to \phi + 2 \pi$, i.e.  from
standard periodic boundary conditions (b.c.)~to b.c.~which are periodic up to a
phase factor $e^{i \theta}$.  There are various ways to realize that in
practice.  If the particle has an electric charge $q$, the introduction of a
magnetic flux $\Phi$ piercing the circle is equivalent to the introduction of a
non-zero $\theta = q \Phi$. The same happens when considering a reference frame
rotating around the axis of the circumference with angular velocity $\omega_p =
\theta / (2 \pi m R^2)$ (see e.g \cite{Krauth_book}).
It should be noted that, analogously to what happens in non-abelian gauge
theories, a non-zero $\theta$ term modifies the Lagrangian by adding a total
derivative to it, therefore it is completely irrelevant at the classical level,
and only plays a role at the quantum level because of the 
non-trivial topology of the configuration space.  

In order to simplify the notation we introduce the variable $x$ defined by
$x=\phi/(2\pi)$ and in the following we will use a system of units such that
$4\pi^2 m R^2=1$. With these conventions the Hamiltonian operator
can be written as 
\begin{equation}\label{eq:H}
\hat{H}=\frac{1}{2}\hat{p}^2+V(x)\ ,
\end{equation}
where $x\in [0,1]$, $V(x)$ is a function of period
$1$ and the wave function $\psi(x)$ satisfies the boundary conditions
$\psi(1)=e^{i\theta}\psi(0)$; the 
corresponding Lagrangian is 
\begin{equation}\label{eq:L}
L[x(t)]=\frac{1}{2}\dot{x}^2-V(x)-\theta\dot{x} \ .
\end{equation}

The partition function $Z(\beta,\theta)$ can be rewritten in the path integral
formalism by means of the Euclidean Lagrangian as ($t\to -i\tilde{t}$)
\begin{equation}
Z(\beta,\theta)=\int\mathscr{D}x(\tilde{t})\exp\left[-\int_0^{\beta}L_E[x(\tilde{t})] 
\mathrm{d}\tilde{t}\right]\ ,
\end{equation}
where the functional integration extends over all functions $x:[0,\beta]\to
[0,1]$ (with the extrema $0$ and $1$ identified) satisfying the boundary
condition $x(0)=x(\beta)$ and the Euclidean Lagrangian $L_E$ is given by
\begin{equation}\label{eq:LE}
L_E[x(\tilde{t})]=\frac{1}{2}\left(\frac{\mathrm{d}x}{\mathrm{d}\tilde{t}}\right)^2
+V(x)+i\theta\frac{\mathrm{d}x}{\mathrm{d}\tilde{t}} \ . 
\end{equation}
Note that the $\theta$ term becomes imaginary after the analytical continuation
to Euclidean time, because it involves a total time derivative; 
this will have important consequences in the following.

Also for generic values of $\theta$ only continuous trajectories have a
non-vanishing weight in the integral, so that paths can be grouped into
homotopy classes labelled by the number of times a path winds around the
circumference while winding once around the Euclidean time. The integral can
thus be decomposed into a sum over the different topologies and an integration
over quantum fluctuations in each fixed topological sector.  As usual in the
context of the semiclassical approximation (see, e.g., Ref.~\cite{Coleman_book}), it
is convenient to select as representatives of the homotopy classes the paths
having the smaller values of the Euclidean action, that correspond to the
solutions of the Euclidean equations of motions.  

This model presents some interesting features even in the non-interacting limit
$V(x) = 0$: in this case the partition function $Z$ at temperature $1/\beta$ is
given in the Hamiltonian formalism by
\begin{equation}\label{eq:Z0}
Z(\beta,\theta)=\sum_{n\in\mathbb{Z}} \exp\left[
-\frac{\beta}{2}(2\pi n+\theta)^2\right]\, .
\end{equation}
The solution of the Euclidean equation of motion having winding number $Q$ is 
\begin{equation}
x_Q(\tilde{t})=\left[\frac{Q}{\beta}\tilde{t}+ \mathrm{const}\right]\,\mathrm{mod}\,1\ ,
\end{equation}
and the corresponding Euclidean action $S_Q=\frac{Q^2}{2\beta}+i\theta Q$
determines the semiclassical exponential weight of the configuration in the
partition function ($S_Q$ is the exponent in Eq.~\eqref{eq:Z0bis}).  When the
action is quadratic in the field the semiclassical approximation is in fact
exact, the integral over the fluctuations at fixed topology is Gaussian and
gives a prefactor of the form $1/\sqrt{\beta}$, so that we finally obtain
from the Lagrangian formalism
\begin{equation}\label{eq:Z0bis}
Z(\beta,\theta)=\frac{1}{\sqrt{2\pi\beta}}\sum_{Q\in\mathbb{Z}}
\exp\left(-\frac{1}{2\beta}Q^2-iQ\theta\right)\ , 
\end{equation}
which could have been also obtained directly from Eq.~(\ref{eq:Z0}) by using
the Poisson summation formula. Eq.~(\ref{eq:Z0bis}) gives a dual representation
of the partition function in Eq.~(\ref{eq:Z0}), in which the high and low $T$
limit are exchanged. Indeed, in Eq.~(\ref{eq:Z0}) only a few terms of the
sum are relevant at low $T$, while at high $T$ all terms contribute and the sum
can be changed into an integral, exactly the opposite happens in the sum over
topological sectors in Eq.~(\ref{eq:Z0bis}).  

From the partition function we can obtain the free energy density
\begin{equation}\label{eq:f}
f(\beta,\theta)=-\frac{1}{\mathcal{V}\beta}\log Z(\beta,\theta)
\end{equation}
that encodes the $\theta$ dependence of the theory. Although for the model
studied in this paper $\mathcal{V}=1$ and there is no need to differentiate
between intensive and extensive quantities, we will use the intensive
lower-case letters to conform to standard notations. Two properties of the free
energy density that are evident in the non-interacting case but are true also
in more general cases are the following: 
\begin{itemize}
\item $f(\beta,\theta)=f(\beta,-\theta)$;
\item $f(\beta,0)\le f(\beta,\theta)$.
\end{itemize}

It is simple to show that the first property is valid if the potential $V(x)$
appearing in the Lagrangian Eq.~\eqref{eq:L} satisfies $V(x)=V(1-x)$ (i.e. if
$V$ is parity invariant): it is sufficient to perform in the path-integral the
change of variable $x(\tilde{t})\to 1-x(\tilde{t})$ and realize that this
corresponds to the substitution $\theta\to -\theta$ in the action.  That
$f(\beta,0)\le f(\beta,\theta)$ is a simple consequence of the fact that in
Euclidean time the $\theta$ term becomes imaginary (see Eq.~\eqref{eq:LE}): the
complex exponential induces cancellations in the path-integral expression of
the partition function and thus $Z(\beta,0)\ge Z(\beta,\theta)$; a fact that,
when the $\theta$ term is interpreted in terms of a non-zero magnetic flux, is
equivalent to a diamagnetic behavior for the charged particle on the
circumference.  Both these properties remain true also in the case of 4d
non-abelian gauge theories and the proof is basically the same; the fact that
the free energy has a minimum at $\theta=0$ is related to the Vafa-Witten
theorem \cite{Vafa:1984xg} and it is at the basis of the Peccei-Quinn solution
of the strong CP problem \cite{Peccei:1977hh, Peccei:1977ur}.

For generic $\beta$ values, the free energy density cannot be written in closed
form using elementary transcendental functions even for the non-interacting
case\footnote{In fact it can be written in terms of the Jacobi $\vartheta_3$
function and the relation between Eqs.~\eqref{eq:Z0},\eqref{eq:Z0bis} in
nothing but the fundamental functional equation for $\vartheta_3$
\cite{Bellman_book, WW_book}.}, however there are two notable limits in which
this is possible: the very low and very high temperature cases. In the extreme
low temperature regime $\beta\gg 1$ we see from Eq.~\eqref{eq:Z0} that for each
$\theta$ value only a single eigenstate contributes and we obtain the
expression
\begin{equation}\label{eq:flargebeta}
f(\beta,\theta)=\frac{1}{2}\,\underset{n\in\mathbb{Z}}{\mathrm{min}}\,
(2\pi n+\theta)^2
\end{equation} 
that is singular at $\theta=\pi$, where a crossing of two energy levels
happens. Large $N$ arguments suggest such a multi-branched $\theta$-dependence
to be present also in 4d $SU(N)$ Yang-Mills theories at zero temperature
\cite{Witten:1980sp, Witten:1998uka}, with a spontaneous breaking of the CP
symmetry taking place at $\theta=\pi$ \cite{Gaiotto:2017yup}; the singularity
at $\theta=\pi$ is present also in QCD for some values of the quark masses
\cite{DiVecchia:1980yfw, Gaiotto:2017tne, DiVecchia:2017xpu}.

In the opposite $\beta\ll 1$ limit, using Eq.~\eqref{eq:Z0bis} we see
that only the modes with $Q=0,\pm 1$ have non-negligible contribution
to the $\theta$ dependence and we obtain
\begin{equation}\label{eq:fsmallbeta}
f(\beta,\theta)=f(\beta,0)-\frac{2}{\beta}e^{-\frac{1}{2\beta}}\cos\theta\ .
\end{equation}
Such an expression is reminiscent of the Dilute Instanton Gas Approximation in
the high temperature phase of 4d non-abelian gauge theories
\cite{Gross:1980br}, however the underlying physics is different, basically
because an underlying spatial volume which diverges in the thermodynamic
limit is missing in the simple 1D model.  In the present model the $\cos\theta$
term is a consequence of the fact that at most a single ``instanton'' is
present, while in the 4d case it is a consequence of the fact that instantons
are almost independent of each other at high temperature. This makes the system
behave ``as if'' a single instanton were  present, although an increasingly
large number of them is in fact present as the thermodynamic limit is
approached.

A parametrization that is commonly used to describe the $\theta$-dependence of
the free energy density in the general case is \cite{Vicari:2008jw}
\begin{equation}\label{eq:fparam}
f(\beta,\theta)=f(\beta,0)+\frac{1}{2}\chi(\beta)\theta^2\left(1+
\sum_{n=1}^{\infty} b_{2n}(\beta)\theta^{2n}\right)\ ,
\end{equation}
where $\chi(\beta)$ is the so-called topological susceptibility and the
coefficients $b_{2n}$ parametrize higher order terms in $\theta$. As previously
noted, in the Euclidean path-integral formulation the $\theta$-term in the
Lagrangian becomes imaginary; this implies that simulations cannot be performed
directly at non-vanishing real $\theta$ values because of a sign problem that
hinders the applicability of standard importance sampling MCMC. 

The most commonly used method to compute the coefficients $\chi$ and $b_{2n}$
appearing in Eq.~\eqref{eq:fparam} is thus to write them in terms of
expectation values computed at $\theta=0$: it is indeed easy to verify that the
first few terms are given by
\begin{equation}\label{eq:def_chi_b2}
\chi=\frac{\langle Q^2\rangle_0}{\beta\mathcal{V}},\quad
b_2=-\frac{\langle Q^4\rangle_{0}-
         3\langle Q^2\rangle^2_{0}}{12\langle Q^2\rangle_{0}} \ , 
\end{equation} 
and in general $b_{2n}$ is proportional to the $2n$-th cumulant of the winding
number distribution at $\theta=0$.  From
Eqs.~\eqref{eq:flargebeta}-\eqref{eq:fsmallbeta} we can see that for the low
and high temperature regimes of the non-interacting theory we have respectively
\begin{equation}\label{eq:low_high_T}
\begin{aligned}
& \beta \gg 1:\quad \chi=1,\quad b_{2n}=0\\
& \beta \ll 1:\quad \chi=\frac{1}{\beta}e^{-\frac{1}{2\beta}}, \quad b_2=-\frac{1}{12},
\quad b_4=\frac{1}{360},\quad \ldots
\end{aligned}
\end{equation}

A different approach that can be used to extract these coefficients is to
perform simulations at imaginary values of the $\theta$ parameter (analytic
continuation method \cite{Azcoiti:2002vk, Alles:2007br, Alles:2014tta,
Panagopoulos:2011rb, DElia:2012pvq, DElia:2013uaf,  Bonati:2015sqt,
Bonati:2016tvi}). Although this approach presents some technical advantages
with respect to the Taylor expansion method, the observed critical slowing down
close to the continuum limit is the same in both approaches, so in this
paper we will concentrate just on simulations performed at $\theta=0$.

\section{Discretization}\label{sec:disc}

The first step to be accomplished in order to perform numerical simulations of
a model by using path-integral techniques is to write down the discretized form
of the Euclidean action. For the case of the Lagrangian Eq.~\eqref{eq:LE} this
is quite a simple task, since it is sufficient to use finite differences
instead of derivatives. The only subtlety stems from the fact that the theory
is defined on a circumference, thus an ambiguity is present in the definition
of the distance between two points. In order to resolve this ambiguity we
introduce the following definition of distance (with sign) on the circumference
of length 1, that correspond to the oriented shortest path between $x$
and $y$:
\begin{equation}
(x-y)\mathrm{mod}\frac{1}{2}=\left\{\begin{array}{lll} 
x-y   & \mathrm{if} & |x-y|\le 1/2 \\ 
x-y-1 & \mathrm{if} & x-y>1/2 \\ 
x-y+1 & \mathrm{if} & x-y<-1/2 
\end{array}\right. \ . 
\label{def:distance}
\end{equation}

With this definition the discretized Euclidean action can be written in the form
\begin{equation}\label{eq:Sdisc}
\begin{aligned}
S=&\frac{1}{2}\sum_j\frac{[(x_{j+1}-x_j)\mathrm{mod}(1/2)]^2}{a} 
+a \sum_j V(x_j)\\
&+i\theta\sum_j [(x_{j+1}-x_j)\mathrm{mod}(1/2)]\ ,
\end{aligned}
\end{equation}
where $x_j\in [0,1)$, $j\in {0,\ldots, N_t-1}$, $a$ is the lattice spacing in
the temporal direction\footnote{Notice that $a$ is a dimensionless parameter.
Indeed, without fixing $\hbar =1$ and $4 \pi^2 m R^2 = 1$ we would have
obtained $a \hbar / (4 \pi^2 m R^2)$ in place of $a$, which is the
dimensionless ratio between the lattice spacing and the typical time scale of
the quantum system. Therefore, with the chosen units, $a \ll 1$ means that the
discretization scale is much smaller than the physical time scale.} and we
will almost always use the thermal boundary conditions $x_{N_t}\equiv x_0$. The
last term in Eq.~\eqref{eq:Sdisc} is the discretizaton of the winding number
$Q$, that in this simple system takes integer values also at non-vanishing
lattice spacing\footnote{Without the proper definition of distance
given in Eq.~(\ref{def:distance}), the discretized topological charge appearing
in Eq.~(\ref{eq:Sdisc}) would be strictly zero.  } (contrary to what happens in
4d gauge theories, see, e.g., Ref. \cite{Vicari:2008jw}).  In the following we
will study the system at $\theta=0$ with the potential
\begin{equation}\label{eq:V}
V(x)=\Omega^2\cos(2\pi x)\ ,
\end{equation}
that for $\Omega=0$ reduces to the non-interacting case, and we will be mainly
interested in the integrated autocorrelation time of the topological
susceptibility, that will be denoted simply by $\tau$.  Another parameter that
is fundamental in order to asses the efficiency of the update methods is
obviously the CPU-time required to perform a single update step.  However, for
the quantum particle moving on a circumference, the computational complexity of
the various update schemes investigated is practically the same, with some
caution needed only for the cases of open boundary conditions and parallel
tempering, as will be discussed later on.

In order to estimate integrated autocorrelation times several procedures exist.
The standard  way, stemming from its very definition,
is to directly integrate the autocorrelation function:
\begin{equation}\label{eq:tauint}
\tau_O=\frac{1}{2}+\sum_{i=1}^{\infty}\frac{\langle O_k O_{k+i}\rangle 
-\langle O\rangle^2}{\langle O^2\rangle-\langle O\rangle^2}\ , 
\end{equation}
where $O$ is a generic primary observable and $O_k$ and 
$O_{k+i}$ denote two generic draws of the observable taken
along the Monte-Carlo simulation at $i$ Markov chain steps apart from
each other. 
The numerical computation of this sum requires some care and standard methods
exist to optimize the integration range, in order to minimize the final error
\cite{MadrasSokal, Wolff:2003sm}. We used the Python implementation described
in \cite{DePalma:2017lww}, that is freely available under the MIT License.

However, probably the most straightforward and practical procedure
is to use the relation 
\cite{Berg_book, MKB} 
\begin{equation}
\delta_{\langle O\rangle}^2 = \frac{2\tau_O}{N_{obs}}
\left(\langle O^2\rangle -\langle O\rangle^2\right) 
= 2 \tau_O\, [\delta_{\langle O\rangle}]_{\mathrm{naive}}^2
\label{eq:taublock}
\end{equation}
where $\delta_{\langle O\rangle}$ is the correct standard error for $\langle
O\rangle$ estimated by properly taking into account autocorrelations, $N_{obs}$
is the size of the sample, and $[\delta_{\langle O\rangle}]_{\mathrm{naive}}$
is the naive standard error computed without taking autocorrelations into
account.  To use this expression we need the value of $\delta_{\langle
O\rangle}$, that can be computed using standard blocking and resampling
procedures (see, e.g., Ref.~\cite{Berg_book, NB_book}).  In all the cases in
which both methods were applicable, we verified that the autocorrelation times
estimated by using the two procedures were compatible with each
other\footnote{Some caution is however needed in the choice of the parameter
$S$ that enters the automatic windowing procedure of \cite{Wolff:2003sm}: this
parameter is typically set to $1.5$ but in some cases it was necessary to use
values up to $15$ for the integral of the autocorrelation function to reach a
plateau around the point automatically chosen by the algorithm (for a different
approach to this problem see \cite{Schaefer:2010hu}).}. Moreover in all
the cases we used time-histories long at least $10^3\tau$ for the largest
values of the autocorrelation time.

In the following section we will describe the results obtained for the
autocorrelation time of the topological susceptibility using different
algorithms. In all these sections, apart from Sec.~\ref{sec:highT}, we will
consider the low temperature regime of the theory, fixing the  temporal extent
of the lattice to the value $aN_t=2$. This corresponds to a temperature
$T=1/2$, which for $\Omega = 0$ and in the chosen units corresponds to $k_B T /
\Delta E = 1 / (2 \pi)^2$, where $\Delta E$ is the energy gap between the
ground state and the first excited state of the free system, so that $T$
is indeed quite small.  We verified to be deep in the low temperature region
for all the values of the parameter $\Omega$ used in our simulations.

\section{Comparison of different approaches}

\subsection{Variation 1: Standard Metropolis}\label{sec:metro}

The first adopted algorithm was the standard Metropolis one \cite{Metropolis}:
we used a 5 hit scheme, with sites updated in lexicographic order and the
proposed update being of the form
\begin{equation}\label{eq:trial}
x\to [x + (1-2r)\Delta]\mathrm{mod}(1),
\end{equation}
where $r\in (0,1)$ is a random number and $\Delta$ is a parameter.  

By increasing the value of $\Delta$ larger variations are proposed, that could
help in reducing autocorrelations, however large values of $\Delta$ have
smaller acceptance rates. In order to investigate the continuum limit of the
autocorrelation time $\tau$ we have to fix the dependence of the parameter
$\Delta$ on the lattice spacing $a$. Given the form of the action in
Eq.~\eqref{eq:Sdisc} it is natural to guess that using $\Delta\propto
\sqrt{a}$ the acceptance rate will be constant as $a\to 0$
\cite{Creutz:1980gp}, and this is indeed what we numerically found.
This is however not \emph{a priori} the optimal choice if our aim is to reduce
the autocorrelation of the topological susceptibility. From some test
runs we concluded that the dependence of $\tau$ on $\Omega$ is practically
negligible and that large values of $\Delta$ correspond to significantly
smaller autocorrelation times, despite the fact that the acceptance rate is
smaller. For this reason we decided to fix the value $\Delta=0.5$ in our
simulation for all the lattice spacings studied.

\begin{figure}[h] 
\centering 
\includegraphics[width=0.9\columnwidth, clip]{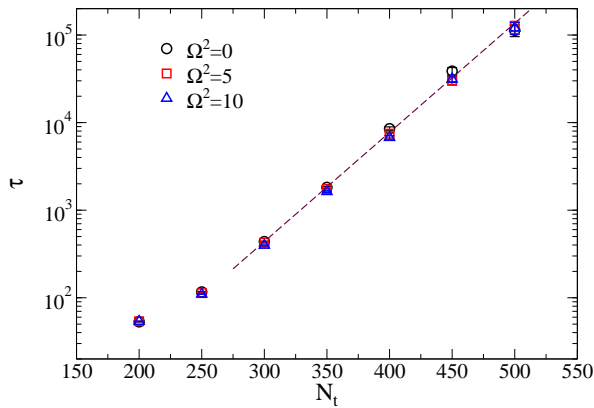}
\caption{Dependence of the autocorrelation time of the topological
susceptibility on the (inverse) lattice spacing for the Metropolis update (with
$\Delta=0.5$ and $aN_t=2$). The dashed line is the result of a fit with the function
$a_0\exp(a_1N_t)$, from which the values $a_0=0.074(10)$ and
$a_1=0.0290(5)$ are obtained.}
\label{fig:metro_autocorr}
\end{figure}

\begin{figure}[t] 
\centering 
\includegraphics[width=0.885\columnwidth, clip]{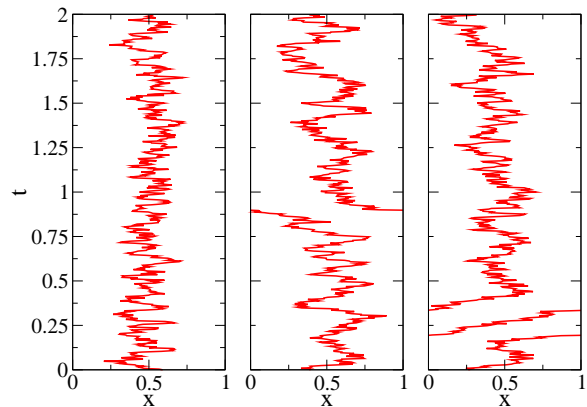}\\
\caption{Examples of configurations corresponding to topological charge $Q=0$
(left), $Q=-1$ (center), $Q=2$ (right) on lattices of temporal extent $N_t=500$
and $T=1/2$. Simulation points are connected by shortest distance lines on the
spatial circumference.}
\label{fig:lattice_instantons}
\end{figure}

In Fig.~\ref{fig:metro_autocorr} we display the scaling of the autocorrelation
time $\tau$ of the topological susceptibility as a function of the inverse of
the lattice spacing (remember that $aN_t=2$) for  simulations performed using
$10^8$ Metropolis updates with $\Delta=0.5$.  The autocorrelation time $\tau$
is well described by an exponential function in $1/a$, which is consistent with
the results obtained in $CP^{N-1}$ models \cite{Campostrini:1992ar,
Flynn:2015uma}, Yang-Mills theories \cite{DelDebbio:2002xa, DelDebbio:2004xh}
and QCD \cite{Alles:1996vn, Schaefer:2010hu, Bonati:2015vqz}.  The critical
slowing down of the winding number emerges, as one approaches the continuum
limit, because for very small lattice spacing the local updates effectively
become continuum diffusive processes. However, in order to change the value of
the winding number by such a type of process, one necessarily has to go across
paths (or configurations, in the case of a field theory) developing a
discontinuity (a cut) somewhere and thus corresponding to very large values of
the action. 

Let us discuss this point more in details: some examples of topologically
nontrivial paths obtained in our MC simulations are shown in
Fig.~\ref{fig:lattice_instantons}. From these figures it is easy to understand
that, in order to change the value of the winding number by a local Metropolis
step, it is necessary to accept an update in which two temporally consecutive
points become very far apart from each other, i.e.  $|x_i-x_{i+1}|\simeq 0.5$.
This is however very unlikely to happen because of the large action of this
configuration. From this very simple argument one can guess the exponent of the
exponential slowing down shown in Fig.~\ref{fig:metro_autocorr} to be given
roughly by
\begin{equation}
\Delta S\simeq \frac{(0.5)^2}{2a}=\frac{1}{4}(0.5)^2N_t=0.0625 N_t\ ,
\end{equation}
which is indeed of the same order of magnitude of the observed value 0.0290(5)
(see Fig.~\ref{fig:metro_autocorr}).

We thus have a consistent picture in which the ``barriers'' that prevent the
changes of the winding number correspond to the very high values of the action
of the configurations one is constrained to go through when moving from one
topological sector to the other by a local algorithm.  Since a shift of the
action of each topological sector as a whole does not touch significantly the
weights of those unlikely configurations, it does not remove the barriers. As a
consequence, it is to be expected that a naive application of the
multicanonical update \cite{Berg:1992qua} or similar approaches (like e.g.
metadynamics~\cite{LaioParr, LaioGervasio}) would not help in removing the
critical slowing down, and this is indeed what we observed.

On the other hand the situation for these classes of update schemes seems to be
better in field theories where one has to deal with non-integer valued
definitions of the winding number in the discretized theory, since a potential
based on a real valued charge can make some distinction, at least in principle,
between different configurations belonging to the same topological sector.
Indeed, very promising results were obtained in Ref.~\cite{Laio:2015era} for 2D
$CP^{N-1}$ models, where the metadynamics was coupled to a discretization of
the winding number that is not strictly integer at non-vanishing lattice
spacing. Encouraging preliminary results obtained using a conceptually related
method, the density of state approach (see e.g. \cite{WangLandau,
Langfeld:2012ah}), have been recently presented in \cite{Cossu:2017sfu}.

\subsection{Variation 2: Tailor method}\label{sec:tailor}

In this section we are going to describe a method that is specifically targeted
at reducing the topological slowing down for the quantum particle moving on a
circumference. The idea is similar, in some sense, to the one used in cluster
updates of spin systems (see, e.g., Ref.~\cite{NB_book}), i.e. to make use of
non-local updates to improve the decorrelation. 

The critical slowing down of the winding number was
linked in the previous section to the necessity of passing through
discontinuities to change the topology when using a local algorithm.
However a non-local update could in principle be able to move from one sector
to the other in one single step, by crossing through the barrier in a sort of
tunnel effect. To remove the slowing down such a step must be able to connect
paths which, while belonging to different topological sectors (typically
adjacent ones), have equal or similar actions.

One can likely invent several kinds of step like that. The one we propose here,
which is specifically devised for the $\Omega = 0$ case, is based on the idea
that if we take a piece of a path and modify it in such a way that $\mathrm{d}x
/\mathrm{d}\tau \to - \mathrm{d}x /\mathrm{d}\tau$ for each $\tau$ (a
reflection around some point will make the job) the action of that piece will
remain the same, while its contribution to the winding will change sign.  If we
can cut away a piece from a starting path, reflect it and sew it back to the
uncut part with a minimal loss of continuity in the cut regions, we will be
able to make a big step in winding number while leaving the action almost
unchanged.

\begin{figure}[t] 
\centering 
\includegraphics[width=0.95\columnwidth, clip]{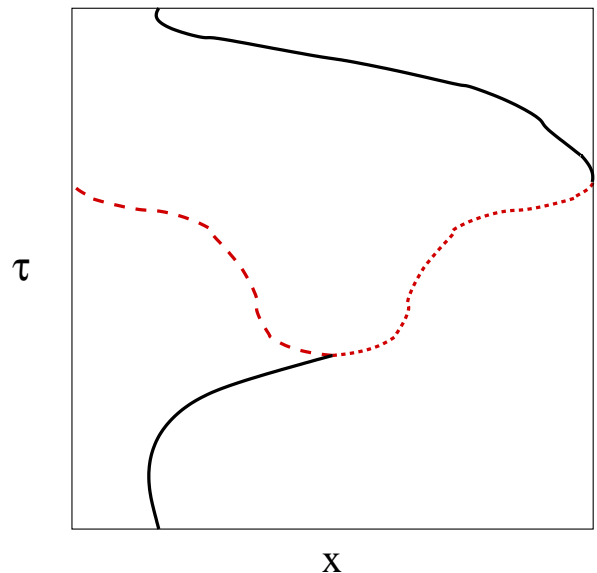}
\caption{An example of the tailor move: a piece of the original path (the
dotted-line one) is cut and sewed back after reversing it (dashed-line one). As
a consequence, a starting path with $Q = 0$ is turned into a path with $Q = -1$
and the same action as before.}
\label{fig:tailor}
\end{figure}

Of course, this sort of path surgery technique will work only for well chosen
paths and cut points. In Fig.~\ref{fig:tailor} we show an example which
clarifies the conditions: the cut piece of path must wind a
half-integer number of times $n/2$ around the circumference, i.e its end points
must be diametrically opposite to each other, so that, after reflection, it can
be sewed back to the uncut path exactly; that also means that $Q$
will change by an integer number, by $-n$ in particular.
Because of this underlying pictorial representation, we have named this
technique as the ``tailor method''.

The above idea can lead to a well defined microcanonical step in the case of
continuous paths. We have now to properly implement it in the case of discrete
paths, and in a way such that detailed balance is satisfied. More in detail,
the algorithm is the following:
\begin{enumerate}
\item randomly choose a value $i_{\mathrm{init}}\in \{0,\ldots N_t-1\}$, to
which the value $x_{i_{\mathrm{init}}}$ of the path is associated. Because of
the periodic boundary condition in the time direction, up to an irrelevant
shift of indices we can assume in the following $i_{\mathrm{init}}=0$.
\item Find the first value $i_{\mathrm{end}}\in \{1,\ldots N_t-1\}$ such that 
\begin{equation}
[x_{i_{\mathrm{end}}}-(x_{0}+0.5)]\mathrm{mod}(1/2)\le \epsilon\ , 
\end{equation}
where $\epsilon$ is a fixed parameter. If no such point exists abort the
update.
\item Propose the update consisting of the change 
\begin{equation}
x_i\to [2x_{0}-x_i]\mathrm{mod}(1/2)
\end{equation}
for all $i$ in $[1,i_{\mathrm{end}}]$.
\item Accept or reject the proposed update with a Metropolis test.
\end{enumerate}

It is easy to see that this algorithm satisfies detailed balance and that, when
the update is accepted, the winding number is changed by $\Delta Q=\pm 1$;
moreover, when $V(x)\equiv 0$, the action of the proposed configuration differs
from the old one only because of the joint at $i_{\mathrm{end}}$, so the
acceptance probability is expected to be reasonably large, at least in the
non-interacting case. Actually, in the continuum limit the acceptance will be
exactly one since the algorithm becomes microcanonical (at least for $\Omega =
0$): in some sense, this kind of update performs at its best right in the
continuum limit. Since this algorithms is almost microcanonical it must
be obviously used together with other more standard algorithms which can
efficiently update the action.  Finally we note that the computational burden
required for such a non-local update is approximately the same as the one
needed to perform a Metropolis sweep for all the points of the lattice.

The algorithm just described could remind the reader of the cluster
algorithms defined by embedding discrete Ising-like variable in a continuum
model \cite{Wolff1989, Brower1989} and, in particular, of the cluster
algorithm described in \cite{Evertz1991} for the solid on solid model of
surface growth; there is however an important difference. The general aim of
these cluster algorithms is to perform non-local updates on large scale
clusters to reduce the autocorrelation times of local observables. In our case
this is not enough and we also have to ensure that during the update the
winding number is changed, which is not guaranteed just by the fact that the
updated region is macroscopically large. In particular we expect the standard
stochastic way of building up the cluster to be less efficient than the
deterministic one explained before to decorrelate topological variables.

The tailor algorithm can be extended without change to the $\Omega \neq 0$ case.  In
this case we expect a loss of performance because of different effects: on one
hand, the action of the path will change after the step, because of the
potential term, so that the acceptance will diminish;  on the other hand, large
values of $\Omega$ will suppress large scale fluctuations of the path, hence
the probability of finding diametrically opposite points. However, in both
cases the problem is not related to the microscopic scale, so that the scaling
of the performance to the continuum limit is expected to be equally good.

\begin{figure}[t] 
\centering 
\includegraphics[width=0.9\columnwidth, clip]{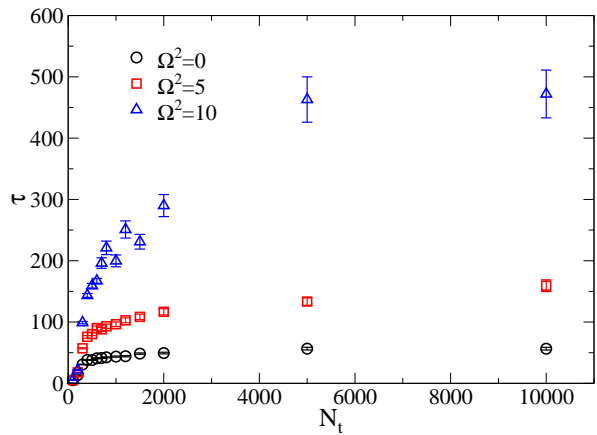}
\caption{Dependence of the autocorrelation time of the topological
susceptibility on the (inverse) lattice spacing when using the tailor
update.}\label{fig:tailor_autocorr}
\end{figure}

In our implementation we used $\epsilon=0.2a$ and a non-local update was
proposed every $10$ Metropolis updates ($5$ hits with $\Delta=0.5$). The
integrated autocorrelation time $\tau$ of the topological susceptibility is
reported in units of the elementary updates in Fig.~\ref{fig:tailor_autocorr},
obtained using a statistics of $10^7$ elementary updates (of which $10^6$ were
non-local tailor updates), from which we can see that the topological slowing
down problem is completely resolved by using the tailor update method.
A nontrivial dependence of the autocorrelation
time on the parameter $\Omega$ is observed, which is easy to understand based
on the discussion above. Indeed for $\Omega^2=0, 5, 10$ the values found for
the acceptance of a tailor move have been respectively $\alpha\simeq
0.5, 0.25, 0.12$. However, the good news are that a sort of saturation of the
autocorrelation time towards the continuum limit is observed in all cases,
independently of the value of $\Omega$.

Given the resolutive nature of this kind of algorithm, it would be a blessing
if an analogous one could be found for field theories with a similar problem of
autocorrelation of topological modes, like QCD. It is important to
stress that for the specific toy model studied in this paper it is surely
possible to find other specific algorithms that drastically reduce or
completely remove the critical slowing down: given the simplicity of the model
it would for example be reasonably simple to implement multilevel/multigrid
algorithm like the ones described in \cite{Ceperley1986, Goodman1986,
Ceperley1995, Goodman1989}, not to mention the possibility of performing
the sampling directly in momentum space. However these methods are known since
quite long times and their applicability to theories involving gauge degrees of
freedom revealed to be very limited. On the contrary, the idea of finding a
global field transformation over a region of spacetime which changes the sign
of the winding density while leaving the action unchanged is easily extendable:
time reversal could for example be used for 4d gauge theories. What is
significantly less easy, in higher dimensions, is to be successful in finding
the region of space-time which can be  cut away and then sewed back after the
field transformation, in such a way that it fits almost perfectly with the
rest, so that the global action change is negligible. For that reason, while
keep thinking about this idea is surely worth doing, let us now turn to more
conventional algorithmic improvements, which are more easily extendable to the
case of field theories.

\subsection{Variation 3: Slab method}\label{sec:slab}

The idea of looking at sub-volumes to estimate the topological susceptibility
is around since some time \cite{Shuryak:1994rr, deForcrand:1998ng} and was
recently suggested as a way out of the freezing problem.  The basic idea is
quite simple and appealing: even if on the whole lattice the winding number
does not change, we can look at local fluctuations and try to extract
information from them.

The way in which this general idea was applied in
Refs.~\cite{Bietenholz:2015rsa, Bietenholz:2016szu} is the following: if we
know the probability distribution $p(Q')_V$ of having a winding number $Q'$ in the
volume $\mathcal{V}$, we can obtain the probability of having winding number
$\tilde{Q}$ in the volume $x\mathcal{V}$ (with $x\in (0,1)$) when the total
winding is frozen to the value $Q$ as
\begin{equation}\label{eq:subq_gen}
p(\tilde{Q})_{x\mathcal{V}}\times p(Q-\tilde{Q})_{(1-x)\mathcal{V}}\ .
\end{equation}
If we now assume that $p(Q)_{\mathcal{V}}\propto \exp(-Q^2/(2\chi
\mathcal{V}))$ it can be shown that, when $Q\equiv 0$ on the whole volume, 
\begin{equation}\label{eq:subq}
\chi_s\equiv \frac{\langle Q^2\rangle_{x\mathcal{V}}}{\mathcal{V}}=\chi x (1-x)
\end{equation} 
and the value of $\chi$ can thus be extracted from sub-volume measurements.

The distribution of the topological charge $Q$ can be quite often approximated,
at least as a first approximation, by a Gaussian behaviour, however this is by
no mean guaranteed. The accuracy of this approximation can be quantitatively
assessed by looking at the coefficients $b_{2n}$ defined in
Eq.~\eqref{eq:fparam}: since they are proportional to the cumulants of the
winding number distribution, they indeed measure how close this distribution is
to a Gaussian. 

For the case of the quantum particle on a circumference we can see from
Eq.~\eqref{eq:low_high_T} that, for the non-interacting case, the distribution
$p(Q)_{\mathcal{V}}$ is well approximated by a continuum Gaussian in the low
temperature regime (it is exactly Gaussian at zero temperature), while this is
no more the case in the high temperature regime. Something very similar happens
also for 4d Yang-Mills theories and QCD: in the low temperature phase the value
of $b_2$ is very small and in fact it goes to zero when the number of color
increases \cite{Witten:1980sp, Witten:1998uka, DiVecchia:1980yfw, 
Bonati:2015sqt, Bonati:2016tvi}, in the high temperature phase the same values
as in Eq.~\eqref{eq:low_high_T} are expected \cite{Gross:1980br, Bonati:2013tt,
Bonati:2015uga, Borsanyi:2015cka, Bonati:2015vqz}, that differs significantly
from the Gaussian ones. 

For the case of the quantum mechanical particle moving on circumference at low
temperature, significant non-gaussianities are expected also in the presence of
a non-vanishing external potential; for this reason in the following of this
section we will restrict to the case $\Omega=0$, in order to use the simple
expression Eq.~\eqref{eq:subq}. It seems reasonable that the formalism can be
extended to take into account also the first few corrections to
non-gaussianity, parametrized, e.g., by $b_2$ and $b_4$, but we will not pursue
this generalization in the present work.

To test the slab method we used lattice spacings small enough that no change of
the winding number was expected to happen in the $10^8$ updates accumulated
(that $Q=0$ in all the cases was also \emph{a posteriori} verified). For the
update we used the same recipe as in the standard Metropolis case, the lattice
temporal extent was fixed once again to $aN_t=2$ and results corresponding to
the different slab sizes (i.e. $x=0.1, \ldots, 0.9$) are completely independent
from each other, since they were extracted from independent runs.  It should be
noted that, while this is not the procedure that would be adopted for
computationally more intensive models, our interest here is just in the
continuum scaling of the autocorrelation time of topological observables
measured at fixed slab size $x$.

\begin{figure}[h] 
\centering 
\includegraphics[width=0.95\columnwidth, clip]{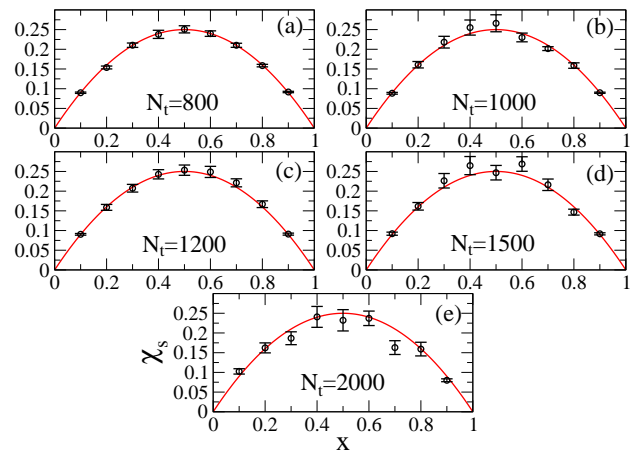}
\caption{Fit to Eq.~\eqref{eq:subq} to extract the topological susceptibility
from sub-volume measurements.}
\label{fig:slab_fit}
\end{figure}

\begin{figure}[h] 
\centering 
\includegraphics[width=0.9\columnwidth, clip]{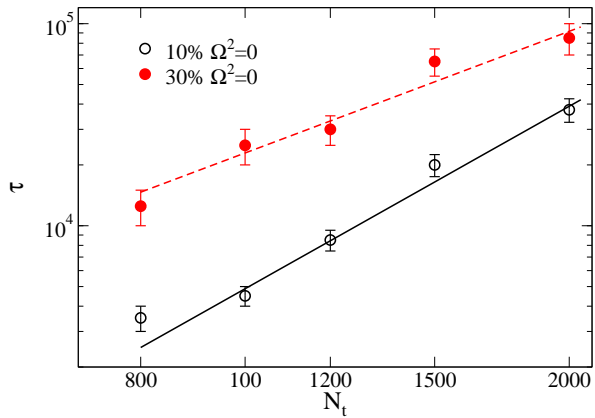}
\caption{Autocorrelation of $\chi_s$, defined in Eq.~\eqref{eq:subq} by using
the sub-volume measures; the case of $x=0.1$ and $0.3$ are shown for
$\Omega=0$. The continuous black line is a fit} of the form $\tau\propto N_t^3$
to the $x=0.1$ data, while the dashed red line is a fit to the $x=0.3$ data
obtained by using the functional form $\tau\propto N_t^2$.
\label{fig:slab_autocorr}
\end{figure}

In Fig.~\ref{fig:slab_fit} we show the fits that have been used to extract
$\chi$ from $\chi_s$. For all $N_t$ values smaller than 2000 the
functional form in Eq.~\eqref{eq:subq} well describes the $x$ dependence of
$\chi_s(x)$ and the values of the topological susceptibility $\chi$ extracted
from the fits are compatible with the expected $\chi=1$ value within errors.
This is not the case for $N_t=2000$, for which we obtained the quite large
$\chi^2/\mathrm{d.o.f}=14/8$ vale and the topological susceptibility estimate
$\chi=0.940(25)$. This could be related to the fact that errors corresponding
to the different $\chi_s(x)$ determinations are for $N_t=2000$ very
inhomogeneous, and a fit to all the $\chi_s(x)$ values at fixed common
statistics is likely not the best choice. Of course this could become a problem
in QCD, where all the slab measures would be extracted not from independent
runs but from the same configurations, thus having always the same statistics
for all the $x$ values. The autocorrelation of the slab measurements is
shown in Fig.~\ref{fig:slab_autocorr} for the case of data corresponding to
$x=0.1$ and $x=0.3$.

The behaviour displayed in Fig.~\ref{fig:slab_autocorr} is quite easy to
understand.  Smaller slabs correspond to smaller autocorrelation times since
they are more sensitive to the local fluctuation of the density of winding
number; note however that, because of the thermal boundary conditions in time
and of the fixed global topology, slices $x$ and $1-x$ are completely
equivalent, the ``worst case'' being thus $x=0.5$. 

Although we do not have data precise enough to draw firm conclusions on the
scaling behaviour of the autocorrelation times in the case of the slab method,
it seems that autocorrelation times at fixed $x$ values are reasonably
described by a power law behaviour in $1/a$, with a power in the range $2\div
3$.  For the case $\Omega=0$ at low temperature it is thus clear that the slab
method improves a lot on the standard Metropolis algorithm. 

How to extend in a systematic way this method to the case in which the
probability distribution of the winding number is non-Gaussian is surely
something worth further investigation. However it has to be remarked that, for
theories that are not defined on a compact space, the probability distribution
of the topological charge always pointwise converges to a Gaussian distribution
in the thermodynamic limit\footnote{This fact is sometimes used in the
literature to argue that the $b_{2n}$ coefficients vanish, however this is not
the case: \emph{all} the cumulants grow with the volume in the thermodynamic
limit and the $b_{2n}$ coefficients stay constant as $V\to\infty$.}. As
a consequence it seems reasonable that the slab method can be safely
applied whenever one is interested just in the topological susceptibility,
provided the volume is large enough (i.e. $\chi V\gg 1$).

\subsection{Variation 4: Open boundary conditions}\label{sec:obc}

The fact that the winding number is an integer number is obviously related, for
the model studied in this paper, to the thermal boundary conditions in time.
Analogously, in 4d non-abelian gauge theories the topological charge is almost
integer on the lattice due to the thermal boundary condition in the time
direction and to the periodic boundary conditions in the space directions, that
are typically used in order to minimize finite size effects.

Given these premises it is natural to think that the critical slowing down of
the topological susceptibility could be alleviated by using different boundary
conditions, that do not constraint the winding number to be integer (or almost
integer). This idea was put forward in \cite{Luscher:2011kk}, where the use of
open boundary condition was suggested, and some interesting variations on the
same theme can be found in \cite{Mages:2015scv, Hasenbusch:2017unr}.

The use of boundary conditions different from the thermal one in the temporal
direction obviously prevent the study of the $T$ dependence of the system.
However we expect the $T=0$ physics to be recovered independently of the
specific boundary condition adopted, if the temporal extent of the lattice is
large enough. In this way we move the topological slowing down problem from the
ultraviolet to the infrared: the winding number is no more discrete and there
can be a winding number density inflow or outflow from the boundary, however we
lose translation invariance and, to avoid contaminations from surface states,
we have to analyze only the part of the lattice that is far from the boundary.

\begin{figure}[h] 
\centering 
\includegraphics[width=0.9\columnwidth, clip]{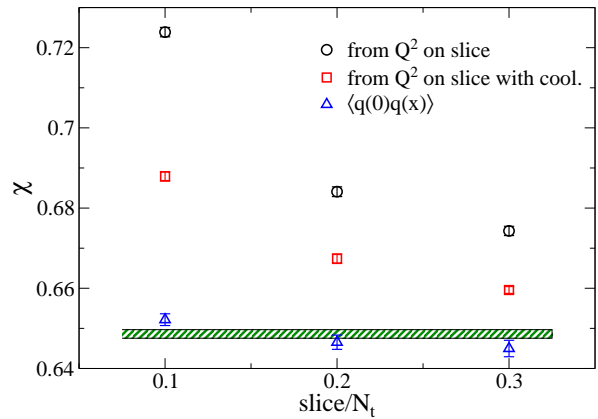}
\caption{Values of the topological susceptibility obtained with different
procedures when open boundary conditions are used: by computing the
susceptibility of the topological charge restricted to the slice, by computing
the susceptibility of the topological charge restricted to the slice after
cooling, by evaluating the integral Eq.~\eqref{eq:qq}. The horizontal band
represents the value obtained from simulation with periodic boundary
conditions.}
\label{fig:test_obc}
\end{figure}

The autocorrelation time of the winding number in the bulk region of the
lattice is thus related to two different phenomena: the inflow/outflow from the
boundary and the diffusion from the boundary to the bulk
\cite{McGlynn:2014bxa}.  Assuming the diffusivity of the topological charge
density to be well defined in the continuum limit, we thus expect the
autocorrelation time of the topological susceptibility in the bulk to scale as
$\tau\propto N_t^2$.

\begin{figure}[h] 
\centering 
\includegraphics[width=0.9\columnwidth, clip]{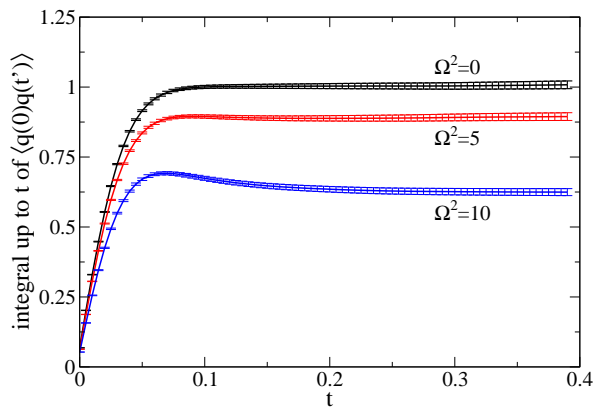}
\caption{An example of the behaviour of $\int_0^t\langle q(0)q(t')\mathrm{d}t'$
as a function of $t$ for three values of $\Omega$, with $t$ in physical units, 
for $1/T=2$ with open boundary conditions and $N_t=400$.}
\label{fig:obc_corr_int}
\end{figure}

When using open boundary conditions, we want to extract information only from
the part of the lattice that is far enough from the boundary as not to be
influenced by its presence; as a consequence the topological susceptibility
cannot be computed simply by using Eq.~\eqref{eq:def_chi_b2}. Making use of the
integral of the topological charge density restricted to a given sub-volume
does not help either: we have to disentangle the true contribution of the
topological susceptibility from the one of the charges that fluctuate across
the boundary of the sub-volume, see Fig.~\ref{fig:test_obc}.  The best way to
compute the topological susceptibility in this case is to write it as the
integral of the two point function of the topological charge density (in the
present case the topological charge density $q$ is simply $\dot{x}(t)$):
\begin{equation}\label{eq:qq}
\chi=\lim_{t\to\infty}\int_{0}^{t}\langle q(0)q(t')\rangle \mathrm{d}t'\ .
\end{equation}
In this equation ``$0$'' is a point in the middle of the lattice, and the lattice
has to be large enough for the asymptotic value of the integral to be reached
before the effects of the boundary become appreciable. 

The integrand of Eq.~\eqref{eq:qq} is however a very singular object for
$t\simeq 0$ \cite{Vicari:1999xx, Seiler:2001je}: even for $\Omega=0$ (at $T=0$)
we have in the continuum $\langle q(0)q(t)\rangle=\delta(t)$. On the other
hand, in the lattice theory, the integral in Eq.~\eqref{eq:qq} reduces to a
finite sum and, for this sum to converge to the integral in the continuum
limit, the integrand must be well behaved as $a\to 0$. The usual procedure that
is used in cases like this is smoothing. 

Many flavours of smoothing exist, like cooling \cite{Berg:1981nw,
Iwasaki:1983bv,Itoh:1984pr, Teper:1985rb, Ilgenfritz:1985dz}, smearing
\cite{Albanese:1987ds, Hasenfratz:2001hp, Morningstar:2003gk} or the gradient
flow \cite{Luscher:2009eq, Luscher:2010iy}, but the main idea is always to
reduce the ultraviolet noise in a local way. While the different variants use
different strategies in order to smooth the configuration (e.g., by taking
local averages or by minimizing the action), in all the cases the net effect is
to introduce a length $\lambda_s$ and to suppress all the field fluctuations on
length-scales smaller than $\lambda_s$.  In order for the lattice sum to
converge to the integral in Eq.~\eqref{eq:qq}, it is important to keep
$\lambda_s$ fixed in physical units as the lattice spacing is reduced.
  
All the smoothing algorithms have been shown to produce compatible results when
their parameters are properly rescaled \cite{Bonati:2014tqa,
Alexandrou:2015yba, Alexandrou:2017hqw, Berg:2016wfw}, so we used the
computationally cheapest procedure: cooling. In order to keep $\lambda_s$
fixed toward the continuum limit we rescaled the number of iterations of the
cooling procedure according to the relation
\begin{equation}\label{eq:nc}
n_c=\mathrm{round}\left[10\left(\frac{0.005}{a}\right)^2\right]\ ,
\end{equation}
where ``round'' denote the rounding to the closest integer. We then \emph{a
posteriori} verified that with this prescription the two point correlator
smoothly converges to its continuum limit, i.e. $\lambda_s$ is indeed fixed in
physical unit.

\begin{figure}[h] 
\centering 
\includegraphics[width=0.9\columnwidth, clip]{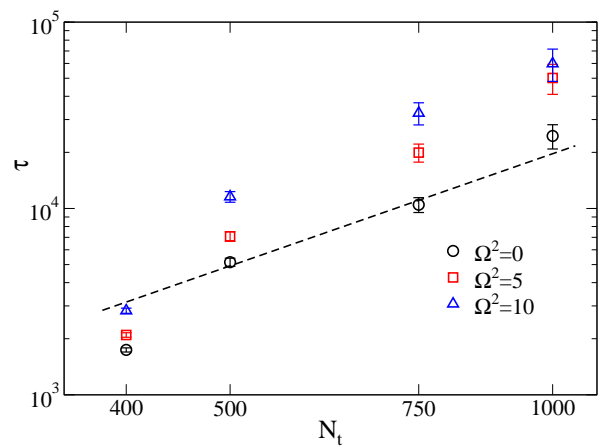}
\caption{Integrated autocorrelation time of the topological susceptibility
obtained from simulations performed with open boundary conditions (with
$aN_t=2$). The dashed line is a fit of the $\Omega=0$ data with the functional
form $a_0N_t^2$.}
\label{fig:obc_auto}
\end{figure}

For the system studied in this paper open boundary conditions in the time
direction can be easily implemented by restricting the first summation in
Eq.~\eqref{eq:Sdisc} to the values $j\in 0,\ldots, N_t-2$; the algorithm used
for the update was the same adopted for the standard Metropolis update.

The typical behaviour of the integral of two point correlation function is
displayed in Fig.~\eqref{fig:obc_corr_int}, from which we see that 
the integral reaches its asymptotic value around $t=0.15, 0.2$ and $0.4$
for $\Omega^2=0, 5$ and $10$ respectively. This implies that a lattice extent 
$aN_t=2$ is large enough also when using open boundary conditions. 

In Fig.~\ref{fig:obc_auto} we show the scaling of the susceptibility
autocorrelation time $\tau$ with the inverse lattice spacing for the three
cases $\Omega^2=0, 5, 10$, obtained from runs consisting of $10^8$
complete Metropolis sweep of the lattice. A fit using the expected
$\tau\propto N_t^2$ behaviour is also displayed, that well describes the data
for small lattice spacing values. The use of the open boundary condition thus
reduces the exponential critical slowing down of periodic boundary conditions
to a quadratic behaviour in the inverse of the lattice spacing. 

This is obviously a great improvement, that however comes together with some
difficulties. For the case of the quantum particle on a circumference the main
difficulty is the impossibility of studying the thermodynamics of the model.
Such a problem is obviously related to the simplicity of the model and it is
not present e.g., for the case of 4d non-abelian gauge theories, in which one
can use the open boundary conditions in the spatial directions. While this
approach should generically work, it could present some difficulties close to a
second order phase transition.

Another source of technical difficulty is related to the definition of the
observables: while for the susceptibility it is relatively easy to use
Eq.~\eqref{eq:qq}, the computation along the same lines of the higher cumulants
of the winding number (needed, e.g., to extract the values of the $b_{2n}$
coefficients) becomes increasingly difficult.

As observed in Sec.~\ref{sec:disc} the autocorrelation time is not the only
figure of merit to be considered to evaluate the effectiveness of an update
scheme, another important one being the CPU-time required to perform a single
update. In particular one is typically interested in minimizing the statistical
error attainable for unit of CPU-time at fixed external parameters (e.g. fixed
lattice size). For the case of open boundary conditions a comparison with 
the standard Metropolis update is thus complicated by two facts:
\begin{itemize}
\item a different estimator of the topological susceptibility is used in the
two cases (i.e.  Eq.~\eqref{eq:def_chi_b2} and Eq.~\eqref{eq:qq})
\item \emph{a priori} different lattice sizes have to be used in the two cases
to have similar finite size effects.
\end{itemize}
For the case studied in this paper these two points turned out not to be
particularly important: the lattices used for the periodic boundary simulations
were large enough to be used without introducing appreciable systematical
errors also for the runs with open boundary conditions; moreover the
variances of the two estimators adopted for the topological susceptibility were
about the same. However, while the critical scaling of the autocorrelation time
should be the same also for other models, these two properties cannot be
expected to hold true in general.

\subsection{Variation 5: Parallel tempering}\label{sec:pt}

Tempering methods have been introduced in the context of the Monte Carlo
simulation of spin glasses, where an exponential critical slowing down is also
present, and they are based on an extended state space approach. This allows
to mix together in a stochastically exact way slowly and quickly decorrelating
simulations.

The tempering approach was applied for the first time to the problem of the
freezing of the topological charge in \cite{Vicari:1992jy}, where simulated
tempering \cite{Marinari:1992qd} was shown to be useful for the case of the 2d
$CP^{N-1}$ model, however a systematic study of its effectiveness has never
been undertaken since then. More recently tempering was used to reduce the
finite size effects associated with the open boundary conditions
\cite{Hasenbusch:2017unr}.

The version of tempering that we used is parallel tempering, introduced in
\cite{SwendsenWang} and later improved in \cite{HukushimaNemoto}. In this
approach several simulations are performed in parallel, each one using
different values of the global parameter that triggers the slowing down, in our
case the lattice spacing.  The global distribution of the whole system is given
by the product of the distributions of the single systems.  For most of the
time the different simulations evolve independently of each other, but
sometimes an exchange is proposed between the configurations corresponding to
different lattice spacing values: of course this implies that the same value of
$N_t$ be used for all copies, so that different lattice spacings will also
correspond to different $N_t a$, i.e.~different temperatures. The exchange of
the configurations $\{x_i\}_{i\in 0,\ldots, N_t-1}$ and $\{x'_i\}_{i\in
0,\ldots,N_t-1}$, corresponding to lattice spacings $a$ and $a'$ respectively,
is accepted or rejected by a Metropolis step with acceptance probability
\begin{equation}
p_{\{x\}\leftrightarrow\{x'\}}=\mathrm{min}\left(1,\frac{e^{-S(\{x\},a')
-S(\{x'\},a)}}{e^{-S(\{x\},a) -S(\{x'\},a')}}\right)\ ,
\end{equation}
where we denoted by $S(\{y\}, \tilde{a})$ the value of the discretized action
in Eq.~\eqref{eq:Sdisc} computed using the configuration $\{y\}$ and the value
$\tilde{a}$ of the lattice spacing (obviously $\theta=0$ for $S$ to be real).
It is easily shown that the algorithm defined in this way satisfies the
detailed balance condition (see, e.g., Refs.~\cite{Berg_book, NB_book}) and the
advantage of the method is that the exchanges drastically reduce the
autocorrelation times, since ``slow'' simulations are speeded up by the
exchanges with the ``fast'' ones.

Let us assume that we are interested in simulating a lattice spacing
$a_{\mathrm{min}}$ at which the standard Metropolis algorithm decorrelates too
slowly. To make use of the parallel tempering approach we have first of all to
select a value $a_{\mathrm{max}}$ at which the Metropolis algorithm is
efficient, then we need to select $N_{PT}$ values of the lattice spacing $a_i$
satisfying
\begin{equation}
a_{\mathrm{min}}\equiv a_0 < a_1 < \cdots < a_{N_{PT}-1}\equiv a_{\mathrm{max}}\ .
\end{equation}
How to select the interpolating values $a_i$ is a non-trivial problem, since
the choice of these values strongly affects the efficiency of the algorithm. 

A criterion that is commonly used to guide the choice of the intermediate
values $a_i$ is the requirement that the probability of the exchange
$a_i\leftrightarrow a_{i+1}$ has to be $i$-independent.  Indeed, one can guess
the autocorrelation time to be approximately given by the time required for a
given configuration to reach the ``quickly decorrelating'' simulation,
decorrelate and come back. If the acceptance probability for the various
exchanges $a_i\leftrightarrow a_{i+1}$ are all equal, the configuration will
perform a random walk between the different copies, and the autocorrelation
will be given by
\begin{equation}\label{eq:tautaumin}
\tau\propto N_{PT}^2\tau_{\mathrm{min}}\ .
\end{equation}
The missing proportionality factor is determined by the acceptance ratio of the
exchange step, which fixes the diffusion constant of the random walk: if
adjacent lattice spacings are chosen far apart from each other, the exchange
will lead to a significant increase of the global action and will be rejected
most of the times.

In general it is not easy to identify intermediate values satisfying the
previous requirements and several procedures exist to optimize their choice
(see, e.g., Ref.~\cite{Katzgraber06, Bittner08, Hasenbusch:2010dm}). For the
case studied in this paper, however, the choice is much simpler: if we consider
for the sake of the simplicity the case $\Omega = 0$, the typical action of a
configuration at lattice spacing $a$ is of the order of $N_t \Delta x^2/a$,
where the typical squared displacement $\Delta x^2$ of the $x$ variable is of
the order of $a$. If we now consider another configuration sampled at lattice
spacing $a'$ and exchange them, the sum of the actions before the exchange will
be of the order of $N_t a /a + N_t a' / a' = 2 N_t$, while after the exchange
it will be of the order of $N_t a'/a + N_t a / a' = N_t (r + 1/r)$  where $r =
a'/a$.  The acceptance probability will thus be $e^{-\Delta}$, with $\Delta$ of
the order of $N_t (r + 1/r - 2)$. As previously noted, for the exchange to be
accepted the two lattice spacings cannot be too different from each other:
$\Delta$ has minimum at $r = 1$ and diverges to positive values for $r \to 0$
or $r \to \infty$. 
This argument is obviously only approximate, since we completely neglected the
role of fluctuations, however it provides an indication that the acceptance
rate has to be a function of $r$ (something that we also verified numerically),
meaning that in order to have equal acceptances one needs to have equal ratios
between adjacent lattice spacings.  The lattice spacing thus plays the same
role as the temperature in a system with temperature independent specific heat,
where it is known that the optimal choice is to use interpolating temperatures
in geometric progression (see, e.g., Ref.~\cite{Pedrescu03}).  As a consequence
we used for the intermediate lattice spacings the values
\begin{equation}\label{eq:geoprog}
a_i=K^ia_{\mathrm{min}},\quad 
K=\left(\frac{a_{\mathrm{max}}}{a_{\mathrm{min}}}\right)^{\frac{1}{N_{PT}-1}} \ .
\end{equation}

To completely fix the update scheme we still have to set the value of $N_{PT}$
in Eq.~\eqref{eq:geoprog}. Of course, one would like to have $N_{PT}$ as small
as possible in order to reduce the number of independent simulations and save
computer time, however, if $N_{PT}$ is too small, the ratio between adjacent
lattice spacings becomes too large and the acceptance probability negligible.
It is also intuitively clear that $N_{PT}$ will have to be larger and larger as
$a_{\mathrm{min}}$ goes to zero with $N_ta_\mathrm{min}$ fixed, i.e. as the
system gets larger and larger in lattice units, since the action is an
extensive quantity. More precisely, it is a standard result (see, e.g.,
Ref.~\cite{HukushimaNemoto}) that, for the acceptance probability of the
exchange $a_i\leftrightarrow a_{i+1}$ to remain constant as $N_t\to\infty$, the
number of copies has to scale asymptotically as follows:
\begin{equation}\label{eq:NPTNt_th}
N_{PT}\propto \sqrt{N_t}\ .
\end{equation}

In fact we found that, for the lattice sizes we used, the previous relation is
not precise enough; for that reason, to keep the exchange acceptance
probability constant as $a_{\mathrm{min}}$ goes to zero we used the empirical
formula
\begin{equation}
\frac{K-1}{\sqrt{a_{\mathrm{min}}}}=C\ ,
\end{equation}
where $K$ is the constant appearing in Eq.~\eqref{eq:geoprog} and the constant
$C$ was fixed to $C=1.4$ after some preliminary tests. Using
Eq.~\eqref{eq:geoprog} we can see that the previous relation implies for
$N_{PT}$ the value
\begin{equation}\label{eq:NPTNt_exp}
N_{PT}=1+\frac{\log(a_{\mathrm{max}}/a_{\mathrm{min}})}{\log(1+C\sqrt{a_{\mathrm{min}}})}\ ,
\end{equation}
that for $a_{\mathrm{min}}\to 0$ (with $a_{\mathrm{max}}$ fixed) is consistent
with Eq.~\eqref{eq:NPTNt_th}.

\begin{figure}[h] 
\centering 
\includegraphics[width=0.9\columnwidth, clip]{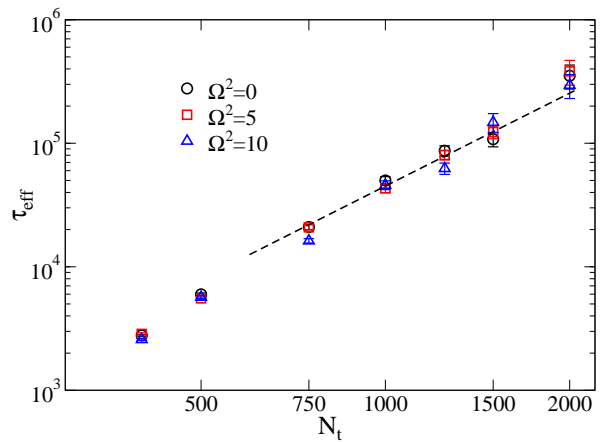}
\caption{Effective integrated autocorrelation time of the topological
susceptibility measured using parallel tempering simulations (with $aN_t=2$).
The dashed line is the result of a fit of the form $\tau_{eff}=a_0N_t^{a_1}$
and the parameter $a_1$ turn out to be $a_1\approx 2.2$.}
\label{fig:pt_auto}
\end{figure}

The largest lattice spacing was fixed in all the runs to
$a_{\mathrm{max}}=0.02$, a value for which the autocorrelation time of the
standard Monte Carlo update is around 10, and an exchange of the configurations
corresponding to neighbouring lattice spacing values was proposed every 20
standard update sweeps. Following Eq.~\eqref{eq:NPTNt_exp} a number of copies
going from 15 to 70 was used depending on the $a_{\mathrm{min}}$ value and, in
order to keep into account the higher computational intensity of a parallel
tempering run with respect to an ordinary Monte Carlo, we introduce the
effective autocorrelation time defined by
\begin{equation}\label{eq:taueff}
\tau_{\mathrm{eff}}=\tau N_{PT}\ ,
\end{equation}
where $\tau$ is the integrated autocorrelation time (in units of the elementary
Metropolis update sweeps) of the topological susceptibility obtained from the
run at smallest lattice spacing of the parallel tempering. 

\begin{figure}[h] 
\centering 
\includegraphics[width=0.9\columnwidth, clip]{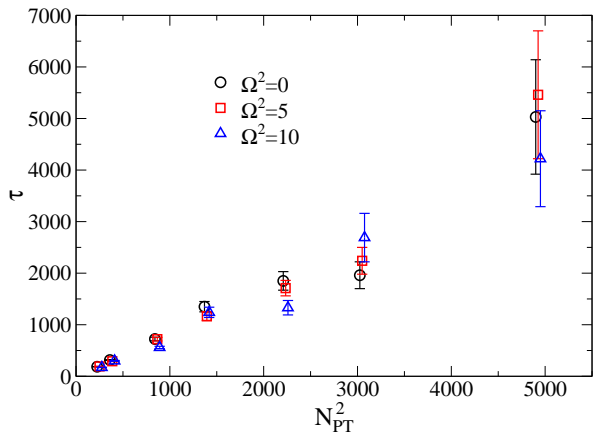}
\caption{Dependence on $N_{PT}$ of the integrated autocorrelation time $\tau$
as measured in the copies with the smaller value of the lattice spacing 
in parallel tempering simulations (with $aN_t=2$). The scaling Eq.~\eqref{eq:tautaumin}
is well satisfied.}
\label{fig:pt_auto_bis}
\end{figure}

The numerical estimates of $\tau_{\mathrm{eff}}$ shown in
Fig.~\eqref{fig:pt_auto} are obtained from simulations with a statistics of
$10^7$ elementary updates (i.e. $5\times 10^5$ parallel tempering exchange
steps) for each replica and again a great improvement with respect to the standard Metropolis
update is clear. The behaviour of the data is roughly compatible with the
scaling $\tau_{\mathrm{eff}}\propto N_t^2$, slightly worse than the expected
$\tau_{\mathrm{eff}}\propto N_t^{3/2}$ that can be obtained from
Eqs.~\eqref{eq:tautaumin}, \eqref{eq:NPTNt_th}, \eqref{eq:taueff}.  This
discrepancy is easily explained by the previous observation that our lattices
are not large enough for the asymptotic expression Eq.~\eqref{eq:NPTNt_th} to
be trusted, indeed the scaling form $\tau\propto N_{PT}^2$ is instead nicely
verified, as shown in Fig.~\ref{fig:pt_auto_bis}.

\subsection{Variation 6: The very high temperature case}\label{sec:highT}

The numerical study of the very high temperature regime presents additional
problems with respect to the low temperature case: the topological
susceptibility approaches zero at high temperature, which means that the
probability $p(Q)$ of observing the value $Q$ of the winding number gets
strongly peaked at $Q=0$. Indeed from Eq.~\eqref{eq:Z0bis} we can see
that in the $\Omega=0$ case
\begin{equation}\label{eq:freeQdistr}
p(Q)=\frac{\exp(-Q^2/(2\beta))}{\sum_{Q'\in\mathbb{Z}}\exp(-Q'^2/(2\beta))}\ ,
\end{equation}
As a consequence it is very difficult to reliably estimate the values of $\chi$
and of the $b_{2n}$ coefficients, since unfeasibly long runs are needed to
compute the momenta $\langle Q^n\rangle$.

For the simple case of the quantum particle moving on a circumference the high
temperature problem is exacerbated by the fact that the physical volume is
fixed from the beginning ($\mathcal{V}=1$ with our conventions). In 4d
non-abelian gauge theories the topological susceptibility also goes to zero in
the high temperature limit \cite{Gross:1980br, Alles:1996nm, Alles:1997qe,
Alles:2000cg, Bonati:2015uga, Berkowitz:2015aua, Bonati:2015vqz}, however the
thermodynamic limit has to be performed. As a consequence, for any fixed value
of $\chi$ (although very small) one can find a volume large enough to observe
with a significant probability states with $Q\neq 0$ (i.e.
$\chi\mathcal{V}\gtrsim 1$).  While this is true in theory, in practice the
size of the lattices is limited by computer resources and as a consequence also
in 4d non-abelian gauge theories the study of the topological properties in the
high temperature phase is particularly problematic.

The practical consequences of this problem are analogous to the ones of the
freezing problem, however it has to be stressed that these two complications
have completely different origins: freezing is a purely algorithmic sampling
problem of the Monte Carlo, while the fact that $p(Q)$ gets strongly peaked at
zero is a physical fact related to the behaviour of the topological
susceptibility.
It is interesting to note that none of the algorithms used so far in this paper
is sufficient to study the high temperature regime, where the lattice spacing
is small enough for the topological freezing to be present and $\chi\ll 1$.  

In order to perform simulations in this regime it is first of all necessary to
enhance the probability of the $Q\neq 0$ states, and the multicanonical
approach is the best suited for this purpose. We thus add 
to the Euclidean action a term corresponding to the potential 
\begin{equation}\label{eq:multicanpot}
V_m(Q)=\left\{ \begin{array}{ll} -\frac{Q^2}{2aN_t\chi_m} & 
     |Q|<Q_{\mathrm{max}}\\
     \rule{0mm}{5mm}-\frac{Q_{\mathrm{max}}^2}{2aN_t\chi_m} & 
     |Q|\ge Q_{\mathrm{max}}\end{array}\right. \ ,
\end{equation}
where $Q_{\mathrm{max}}$ and $\chi_m$ are parameters of the algorithm and this
potential will enter the reweight analysis procedure.  The adopted values for these
parameters are $Q_{\mathrm{max}}=5$ and $\chi_m=1$ (this is the
optimal value for $\Omega=0$ according to Eq.~\eqref{eq:freeQdistr}) and
simulations have been performed using the same procedure as for the simple
Metropolis update. No particular optimization of the simulation parameters has
been investigated, since in this section we are more interested in a proof of
principle about the feasibility, rather than in an actual optimization of the
runs. 

As previously noted the use of the multicanonical ensemble does not
solve the freezing problem, so we have to use some other algorithmic
improvement on top of the multicanonical approach.  The tailor method cannot be
used for this purpose, since the typical $Q=0$ configurations of the
path-integral Monte-Carlo are almost straight at high temperature, also when
the multicanonical term Eq.~\eqref{eq:multicanpot} is present in the Euclidean
action.  As a consequence it is very unlikely to find the two diametrically
opposite points that are needed for the tailor update to succeed.

\begin{figure}[h] 
\centering 
\includegraphics[width=0.9\columnwidth, clip]{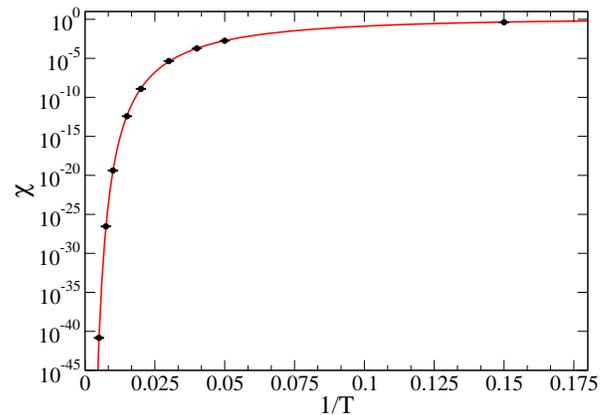}
\caption{Topological susceptibility in the very high temperature case. Just the
$\Omega=0$ data are plotted since the difference with the $\Omega\neq 0$ data
is invisible on this scale. The red line is the theoretical expectation of
Eq.~\eqref{eq:low_high_T}.}
\label{fig:hight}
\end{figure}

In order to avoid the freezing problem we thus adopted parallel tempering
in combination with multicanonical sampling. To verify that this
approach indeed allows to efficiently perform simulations in the high
temperature region we used a lattice with temporal extent $N_t=200$ and lattice
spacings in the range $[7.5\times 10^{-4}, 2.5\times 10^{-5}]$, corresponding
to $T\in [7, 200]$. We did not perform any optimization of the parameters
entering in the parallel tempering, and in all the cases we just used
$a_{\mathrm{max}}=0.05$ and $N_{PT}=80$. 

The results obtained for the topological susceptibility (with a
statistics of $10^7$ elementary updates for each replica) using this approach
are shown in Fig.~\ref{fig:hight} together with the theoretical expectation for
the non-interacting case Eq.~\eqref{eq:low_high_T}. The presence of a
non-vanishing $\Omega$ is in this case completely irrelevant and numerical
simulations correctly reproduce the theoretical predictions over 40 orders of
magnitude.  Although no specific optimization of the simulation parameters has
been pursued, in all cases $\tau_{\mathrm{eff}}$ was in the range $10^4\div
10^5$, with a very slow increase as the lattice spacing is decreased.  For the
copies with the largest lattice spacings (corresponding to the lowest
temperatures) we tried to also add in the simulation a tailor update step,
whose effect was to reduced the autocorrelation times by an additional factor
$2\div 3$. 

\begin{figure}[t] 
\centering 
\includegraphics[width=0.9\columnwidth, clip]{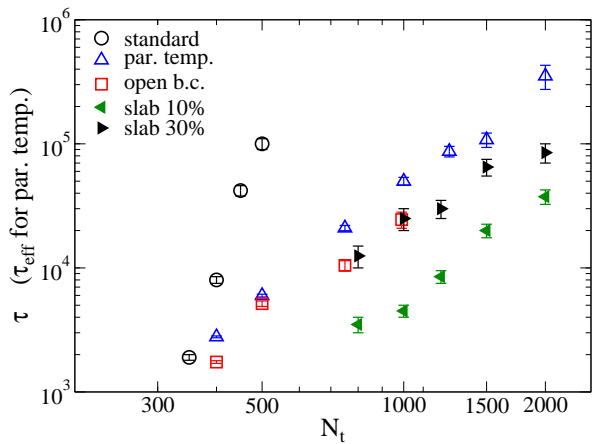}
\caption{Comparison of the results obtained by using different simulation
algorithms in the low temperature phase for $\Omega=0$. Results obtained
by the Tailor update are not shown in order to make the figure more readable.}
\label{fig:compare0}
\end{figure}

\section{Conclusions}\label{sec:concl}

In this study, we have considered a quantum mechanical problem which furnishes
one of the simplest examples of path integral characterized by a topological
classification of the configurations, and by the possible introduction of a
topological $\theta$ term.  After reviewing its main properties and some
interesting exact duality relations connecting the high temperature and the low
temperature regime, our main focus has been on the numerical investigation of
the discretized path integral by Monte-Carlo simulations.

\begin{figure}[t] 
\centering 
\includegraphics[width=0.9\columnwidth, clip]{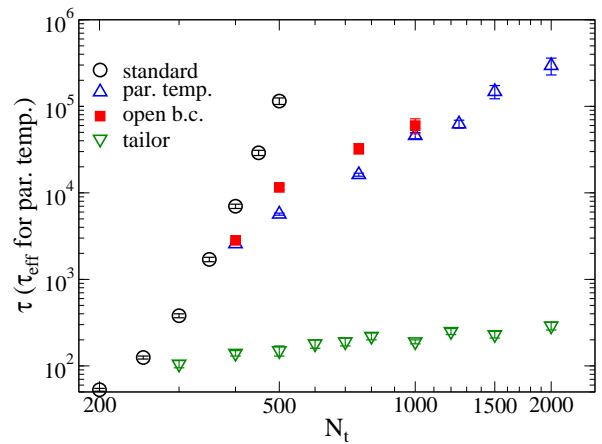}
\caption{Comparison of the results obtained by using different simulation 
algorithms in the low temperature phase for $\Omega^2=10$.}
\label{fig:compare10}
\end{figure}

Indeed, a hard algorithmic problem which this humble model shares with nobler
examples, like QCD, is the emergence of a critical slowing down of topological
modes as the continuum limit is approached, leading eventually to the
impossibility of properly sampling the path integral.  For that reason, we have
decided to explore several algorithmic improvements, some of them already
proposed in the context of QCD and other field theories, in order to perform a
systematic and comparative investigation of them. A summary of our results,
obtained both for the free case and in the presence of a $\cos$-like potential,
is reported in Figs.~\ref{fig:compare0} and \ref{fig:compare10}.

When investigating a simplified toy model in place of a more complex theory,
one always faces the risk of saying something which is relevant just to the toy
model and not to the original problem. We have not spent much efforts in
avoiding this risk, indeed the best algorithm we have developed for the toy
model is what we have named the ``tailor method'', which by means of clever
cuts and seams of selected pieces of configurations, achieves a direct
tunneling between adjacent topological sectors which eliminates completely the
problem of the so-called topological barriers. We have argued that, while the
underlying symmetry concepts at the basis of the method could be exported to
more complex cases, finding the proper cuts and seams is a significantly harder
task in more than one space-time dimension.

However, we think that some of our findings could indeed be quite relevant also
for theories like QCD. At $T = 0$ the slab method, the method of open boundary
conditions and parallel tempering provide the best results, achieving
comparable efficiency in reducing the autocorrelation time of the
topological susceptibility, greatly improving on the basic Metropolis scheme.
From our results we obviously can not draw firm conclusions on which of these
algorithms would best perform in a realistic simulation of a Yang-Mills theory
or QCD, since non-universal features (like e.g. the prefactor of the
scaling laws of $\tau$) could be significantly different, and the relative
computational weight of the different approaches can become different from
one. However parallel tempering appears to be the most flexible method of
this group: the slab method is (at least in its present form) limited to the
case of a Gaussian probability distribution, while the use of open boundary
conditions could create some technical difficulties close to a second order
phase transition and in the computation of the $b_{2n}$ coefficients.

The situation is quite different when one considers the high-$T$ regime.  In
this case one has a new hard problem, in addition to the freezing one, namely
the exponential suppression of the weight of topological sectors with $Q \neq
0$, which therefore require exponentially large statistics in order to be
properly sampled.  Some of the methods explored at low $T$, like the slab
method and open boundary conditions, are not available in this case for
the model studied here, because of the short time direction and of the absence
of other space-time dimensions.  The optimal strategy we have found to defeat
both problems at the same time is to make use of parallel tempering in the
context of a multicanonical simulation (metadynamics would work equally well).
We believe that this could be a good suggestion to approach the equivalent
problem one has to face in exploring the topological properties of QCD in the
very high temperature regime, something which is very relevant in the context
of axion phenomenology~\cite{Berkowitz:2015aua, Borsanyi:2015cka,
Kitano:2015fla, Trunin:2015yda, Bonati:2015vqz, Petreczky:2016vrs,
Frison:2016vuc, Borsanyi:2016ksw}.

\emph{Acknowledgement} We thank Francesco Sanfilippo for many useful
discussions. CB wishes to thank Barbara De Palma for clarifying discussions.


%

\end{document}